\documentclass[10pt,journal,compsoc]{IEEEtran}
\ifCLASSOPTIONcompsoc
  \usepackage[nocompress]{cite}
\else
  \usepackage{cite}
\fi
\usepackage{graphicx}
\usepackage{float}

\usepackage{subcaption}

\ifCLASSINFOpdf
\else
\fi
\begin{document}
%
\title{Speeding up Block Propagation in Blockchain Network: Uncoded and Coded Designs}
%
%
%
%

\author{Lihao~Zhang,
        Taotao~Wang,~\IEEEmembership{Member,~IEEE,}
        and~Soung~Chang~Liew,~\IEEEmembership{Fellow,~IEEE}
\IEEEcompsocitemizethanks{\IEEEcompsocthanksitem Lihao Zhang and Soung Chang Liew are with the Department
of Information Engineering, ,The Chinese University of Hong Kong, Hong Kong.\protect\\
E-mail: zl018@ie.cuhk.edu.hk, soung@ie.cuhk.edu.hk
\IEEEcompsocthanksitem Taotao Wang is with the College of Electronics and Information Engineering, Shenzhen University, Shenzhen 518060, China.
E-mail: ttwang@szu.edu.cn}
\thanks{Manuscript received xxx, xxx; revised xxx, xxx.}}

%
%

\markboth{Journal of \LaTeX\ Class Files,~Vol.~xx, No.~x, December~2020}%
{Shell \MakeLowercase{\textit{et al.}}: Speeding up Block Propagation in Bitcoin Network: Uncoded and Coded Designs}
%



\IEEEtitleabstractindextext{%
\begin{abstract}
We design and validate new block propagation protocols for the peer-to-peer (P2P) network of the Bitcoin blockchain. Despite its strong protection for security and privacy, the current Bitcoin blockchain can only support a low number of transactions per second (TPS). In this work, we redesign the current Bitcoin’s networking protocol to increase TPS without changing vital components in its consensus-building protocol. In particular, we improve the compact-block relaying protocol to enable the propagation of blocks containing a massive number of transactions without inducing extra propagation latencies. Our improvements consist of (i) replacing the existing store-and-forward compact-block relaying scheme with a cut-through compact-block relaying scheme; (ii) exploiting rateless erasure codes for P2P networks to increase block-propagation efficiency. Since our protocols only need to rework the current Bitcoin’s networking protocol and does not modify the data structures and crypto-functional components, they can be seamlessly incorporated into the existing Bitcoin blockchain. To validate our designs, we perform analysis on our protocols and implement a Bitcoin network simulator on NS3 to run different block propagation protocols. The analysis and experimental results confirm that our new block propagation protocols could increase the TPS of the Bitcoin blockchain by 100x without compromising security and consensus-building. 
\end{abstract}

}

\maketitle{}

\IEEEdisplaynontitleabstractindextext

%
\IEEEpeerreviewmaketitle

\IEEEraisesectionheading{\section{Introduction}\label{sec:introduction}}

%
%
%
%
\IEEEPARstart{B}{lockchain} was proposed as a supporting technology for Bitcoin \cite{nakamoto2019bitcoin}, the first decentralized cryptocurrency. After Bitcoin, other decentralized cryptocurrencies (e.g., Litecoin \cite {reed2017litecoin}, Ethereum \cite{wood2014ethereum}) quickly emerged. The blockchains of these cryptocurrencies use the Nakamoto's proof-of-work (PoW) protocol to build consensus among distributed nodes. Blockchain has, by now, become a cutting-edge technology in the fields of FinTech, Internet of Things (IoT), and supply chains \cite{du2019affordances,kshetri2017can,kouhizadeh2018blockchain}, thanks to its ability to enable Byzantine agreement over a permission-less decentralized network \cite{singh2016blockchain}. 

A weakness of the current blockchains is the low on-chain transaction throughput. For example, the throughput of Bitcoin is around 5 $\sim$ 7 transactions per second (TPS), and that of Ethereum is around 40 TPS \cite{gobel2017increased}. Both are extremely low compared to around 110 TPS of PayPal and 1700 TPS of Visa. Its low transaction throughput hampers the widespread adoption of today's blockchain technology. A straightforward method to increase TPS is to enlarge the block size so that a block can carry more transactions. However, the propagation of large blocks in the network may incur huge delays that compromise the blockchains' security and integrity \cite{gervais2016security}, and thus it is not a good idea to increase the TPS by merely increasing the block size. Consequently, new consensus-building protocols and new specially deployed networking infrastructures have been proposed as solutions to increase the TPS of blockchains (see discussions of these related works in Section \ref{section:relatedWork}).  

In this work, we put forth a new block propagation protocol to propagate large blocks containing a large number of transactions without increasing the block relay delay. Unlike the existing solution that requires changing the consensus-building protocol or deploying new network infrastructures, we build our block propagation protocol upon the compact-block relaying protocol that is already adopted by the current Bitcoin network \cite{corallo2017compact}. Compact-block relaying reduces the block relay delay by compressing blocks that contain transactions (around 250 Bytes each in Bitcoin) into compact blocks that contain transaction hashes (6 Bytes each). Therefore, compact-block relaying can include more transactions into each compact block while maintaining the same relay delay, thus increasing the TPS without compromising the blockchain's security. 

In this paper, we further boost TPS by further increasing compact-block size including even more transaction hashes into each compact block. However, simply increasing compact-block size induces extra propagation delays that may compromise blockchain's security. We devise methods to keep the propagation delays at bay while increasing the compact-block size. We adopt a two-pronged approach: 1) we replace the store-and-forward compact-block relaying scheme with a cut-through forwarding scheme; 2) we apply rateless erasure codes to increase the efficiency of block propagation. The contributions of this work are listed as follows.
\begin{enumerate}
\item We put forth a new block propagation protocol that replaces the store-and-forward compact-block relaying scheme with a cut-through forwarding scheme. Our new cut-through compact-block relaying scheme can propagate large compact blocks without inducing extra relay. The original compact-block relaying \cite{corallo2017compact,ozisik2017graphene} is a store-and-forward scheme in which a whole compact block must be received before it is forwarded. With cut-through forwarding, a node receives small chunks of a compact block while forwarding earlier-received chunks, allowing reception and forwarding of a compact block to progress in parallel.
\item We apply rateless erasure codes to compact blocks. Rateless erasure codes allow the recovery of the source symbols using a subset of the encoded symbols. Instead of distributing the original compact block’s source symbols, the compact block's source distributes the compact block’s encoded symbols. Importantly, rateless erasure codes allow a peer to construct source symbols by retrieving the encoded symbols from multiple peers. Our coded design benefits from the peer-to-peer (P2P) network topology of the Bitcoin blockchain by efficiently utilizing the upload bandwidths of all peers \cite{gkantsidis2005network}.
\item We perform a theoretical analysis of our protocols assuming a simple linear network. The analysis results confirm that our protocols can obtain significant TPS gain. Moreover, to evaluate our protocols in a practical network, we implemented a Bitcoin network simulator. We simulated propagating huge compact blocks containing a large number of transaction hashes to improve TPS. Our results indicate that our block propagation protocols can increase TPS by 100x while maintaining the same propagation delay as the conventional block propagation protocols. 
\end{enumerate}	
Since our design is built upon compact-block relaying that has been implemented into the Bitcoin network, we believe our design can be readily deployed in the existing blockchain networks. 

The rest of this paper is organized as follows. Section \ref{section:relatedWork} discusses related works. Section \ref{section:background} presents the background of blockchain. Section \ref{section:Uncoded} and Section \ref{section:Coded} introduce our uncoded and coded designs. Section \ref{section:analysis} analyzes the performance of our two-pronged approach. Section \ref{section:simulation} discusses our experimental results, and Section \ref{section:conclusion} concludes this work.

\section{Related Work}\label{section:relatedWork}
To improve the TPS of the Bitcoin blockchain, \cite{consensus2019, natoli2019deconstructing} changed the Nakamoto's PoW consensus protocol . These clean-slate designs modified data structures and many crypto functional components in the Bitcoin blockchain’s consensus protocol. Consequently, these new consensus protocols are incompatible with the today’s Bitcoin blockchain. By contrast, our work aims to increase TPS by redesigning the networking protocol without changing other key functional components. Furthermore, our work does not require the building of new network infrastructures. There has been little prior work along this line. 

Ref. \cite{mccorry2020sok} discussed the shortcoming of the networking aspects of the Bitcoin blockchain. To reduce block propagation delay, \cite{corallo2017compact, ozisik2017graphene} devised compact-block relaying that transmits a compressed block consisting of the hashes of transactions in the block; transactions are transmitted (upon feedback from the receiver) only if they are missing at the receiver. Investigation in \cite{decker2013information} preemptively announce a block's availability before the complete reception of the whole block. Exploiting the multicasting capability of network nodes in blockchain, \cite{chawla2019velocity} employs fountain codes \cite{luby2002lt} to enable a full node to obtain a block from multiple peers. However, \cite{chawla2019velocity} still relies on store-and-forwarding rather than cut-through forwarding. 

Several works advocated building new network infrastructures for speeding up the block propagation of the Bitcoin blockchain. For example, \cite{klarman2018bloxroute} used a blockchain distribution network (BDN) with high throughput servers to speed up the propagation of large blocks. In particular, \cite{klarman2018bloxroute} adopted cut-through forwarding within the BDN, but not at distributed blockchain nodes. The BDN approach requires building a new infrastructure controlled by a central authority, partially offsetting decentralized blockchain systems' many advantages. FIBRE (Fast Internet Bitcoin Relay Engine) in \cite{corallo2017fibre} is a transport-layer protocol that uses UDP with forward error correction to decrease the delays caused by packet loss. It also introduces data compression to reduce the amount of network traffic. Our work, focusing on the blockchain networking protocol \footnote{In the TCP/IP and OSI nomenclature, the blockchain networking protocol resides in the application layer. In this paper, we do not change IP or TCP protocol at all.}, is fully compatible with FIBRE and can be merged with it.

The idea of using cut-through forwarding to improve block propagation is also employed by Falcon \cite{falcon}. The differences between Falcon and our work here are as follows: i) Falcon implements cut-through forwarding on specially deployed relay nodes rather than on the existing blockchain nodes; ii) Falcon is a commercial project and there lacks performance analysis on its block propagation; iii) Falcon does not employ rateless erasure codes to increase the efficiency of block propagation. 

\section{Bitcoin Blockchain Background}\label{section:background}
This section reviews the background for the Bitcoin blockchain, including its data structure, PoW consensus protocol, network topology, and block propagation protocols. After that, we present our compact block relaying scheme. 

\subsection{Data Structure of Blockchain}
Blockchain is a decentralized append-only ledger for digital assets. A blockchain is replicated and shared among participants. The transactions in the system are contained in concatenated blocks in the blockchain. A block contains a header and multiple transactions. The state of a blockchain maintained by a participant changes according to the blocks in the blockchain. 

We can write a block as $B = \left[ {H,{T_1},{T_2}, \cdots ,{T_K}} \right]$ where $H$ denotes the header, $\left\{ {{T_1},{T_2}, \cdots ,{T_K}} \right\}$ denotes the transactions in the block, and $K$ denotes the number of transactions in a block. The header $H$ contains the hash computed from the content of the preceding block, the Merkle root of the transactions in this block, a nonce generated by the PoW consensus protocol and a number indicating the mining target. Each block must refer to its preceding block by placing the hash of its preceding block in its header, and the blocks form a chain of blocks arranged in chronological order. 

\subsection{PoW Consensus Protocol}
A consensus protocol coordinates the blockchain's updates to ensure chronological ordering of the blocks and to ensure the blockchain's integrity and consistency across geographically distributed nodes. Bitcoin \cite{nakamoto2019bitcoin}, as the first implementation of blockchain, introduces the Proof-of-Work (PoW) consensus protocol. Tens of thousands of distributed nodes adopt the PoW consensus protocol to achieve data consistency (i.e., ensuring the blockchains maintained by them are the same). Before adding blocks to the blockchain, a Bitcoin node has to prove that it has performed some work known as PoW. In essence, a node must find a nonce input to a hash function so that the hash value is less than a target number, as expressed by
\begin{equation}
\label{eqn:mininghash}
h\left( {n,p,m} \right) < D
\end{equation}
where $n$ is the nonce, $p$ is the hash of the preceding block, $m$ is the Merkle root of the transactions in the block, $h\left(  \cdot  \right)$ is a hash function, and $D$ is the mining target that is small with respect to the whole range of possible hash function outputs. The target $D$ is determined by a difficulty level set by the Bitcoin network. The header $H$ of a block contains $n$, $p$, $m$ and $D$. The difficulty level is dynamically tuned by the Bitcoin protocol, which ensures that the participating nodes, as a whole, produce an average of one block every ten minutes. The process of solving the PoW puzzle is called mining, and the nodes that perform the mining function are known as miners.

When other nodes receive a new block broadcasted from the miner, they verify the block locally and independently. The verification of a block can be divided into two parts: 1) the verification of the PoW solution, i.e., verifying whether the nonce $n$ contained in the block header fulfills (\ref{eqn:mininghash}) given the other block header's contents; 2) the verification of the transactions contained in the block body, i.e., verifying the validity of each transaction and verifying whether the Merkle root of all the transactions included in the block body is consistent with the Merkle root contained in the block header. If block passes verification, this block is appended to the local blockchain of the node. 

\subsection{Blockchain Network Topology}
The network of the Bitcoin blockchain is based on an unstructured P2P network. The Bitcoin Core project \cite{bitcoincore} implements the Bitcoin networking protocol. When a node initializes, it attempts to discover a set of peers to establish outgoing or incoming\footnote{Outgoing connections are initiated by the node itself, and incoming connections are initiated by other nodes. When a Bitcoin node boots up, it asks the DNS seed nodes for a list of Bitcoin nodes’ IP addresses. Then, it selects a subset of these addresses and initiates up to 8 outgoing connections. Meanwhile, this node’s IP address will be logged and, going forward, sent to other nodes by the DNS seed nodes. This node will accept up to 117 incoming connections. Super nodes may establish more than 8 outgoing connections.} TCP connections. These connections are used for transaction and block propagation. Each node maintains a list of peers' IP addresses. According to the default protocol in the Bitcoin Core client, a node in the Bitcoin network initiates up to 8 outgoing connections and accepts up to 117 incoming connections. While the Bitcoin infrastructure does not support the discovery of the overall network topology — this is to secure the network from potential network attacks such as Eclipse\cite{heilman2015eclipse}, Sybil\cite{bissias2014sybil} — there have been works that acquire the network topology using the information in the Bitcoin blockchain protocol messages. A Bitcoin P2P Network Sniffer\cite{donet2014bitcoin} can connect to a Bitcoin node and listen to network events such as block broadcast or transaction broadcast. The collected data can be used to infer the size of the Bitcoin P2P network and the geographic distributions of nodes. We use this type of network topology information to model our blockchain network simulator, as discussed in Section \ref{section:simulation}.

\subsection{Transaction and Block Dissemination}
The network that supports the dissemination of transactions and blocks in the blockchain is a P2P network overlaid over the Internet. Each node keeps a replica of the complete blockchain. There is no central coordinator to ensure the consistency of the replicas across the different nodes. Rather, the nodes use a simple gossip protocol to propagate messages containing transactions and blocks to update and synchronize their ledger replicas. 

\begin{figure*}[!t]
\centering
\includegraphics[width=5in]{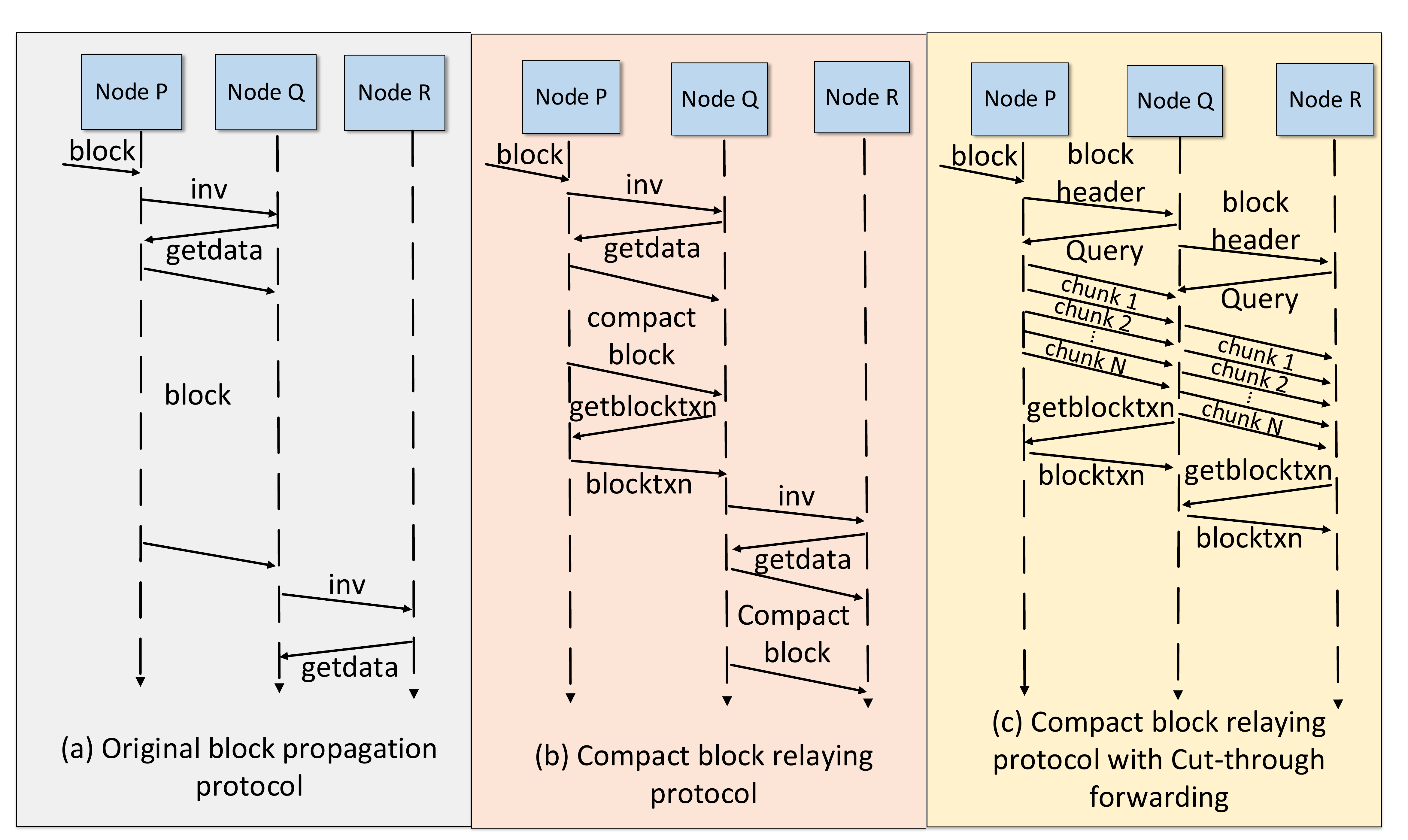}
\caption{The forwarding of a block over two hops by three different block propagation protocols: (a) the original block propagation protocol; (b) the compact block relaying protocol; (c) the cut-through compact block relaying.}
\label{fig_1}
\end{figure*}

Let us first describe the block propagation protocol of standard relaying (SR) as illustrated in Fig. \ref{fig_1}(a). To avoid unnecessary block forwarding, a peer does not forward blocks to its peer immediately. Instead, an inv \cite{decker2013information} containing a summary of the blocks available—specifically, hashes of the blocks—is sent to a neighbor (node B in Fig. \ref{fig_1}(a)). When a node receives an inv message, it issues a getdata message to the sender, only requesting for blocks that it does not have. The SR protocol, although simple, is inefficient and incurs large propagation delay and has low transaction throughput.

Compact-block relaying (CBR) \cite{corallo2017compact}, as illustrated in Fig. \ref{fig_1}(b), is a solution to decrease propagation delay. CBR also announces blocks available at the sender through an inv. However, upon receiving a getdata message, the sender only returns a compact block containing a summary of the transactions included in the entire block (i.e., the hashes of the transactions). The receiver then only requests for the transactions in the block that it does not have (blocktxn in Fig. \ref{fig_1}(b)) rather than the whole block (block in Fig. \ref{fig_1}(a)). Since blocktxn contains only a fraction of all transactions in a block, the bandwidth needed for data propagation is reduced. Although CBR decreases the propagation delay, the sender still needs to have the whole compact block in hand before relaying it. This relaying is called "stored-and-forward". While store-and-forward relaying works well for small compact blocks, if we intend to increase the block size to boost TPS, the compact blocks become large themselves. When relaying compact blocks of large size, the store-and-forward CBR can incur large latency. 

%
%

\section{Uncoded Design: Compact-Block Relaying with Cut-through Forwarding}\label{section:Uncoded}
This section presents the design of a new CBR protocol. Although the original store-and-forward CBR can increase the TPS to some extent, according to our experiment result (see Section \ref{section:simulation}.2), using the store-and-forward CBR to propagate compact blocks of large size can lead to large propagation delays.  

As illustrated in Fig. \ref{fig_1}(c), the main essence of our new CBR protocol is to allow parallel relaying of large compact blocks. A node P forwards a compact block to another node Q in small chunks. Node Q then forwards the earlier chunks to node R as later chunks are received. In particular, node Q does not wait for all the chunks of the compact block to be received before forwarding the earlier received chunks. We refer to our block propagation protocol as cut-through compact-block relaying (Cut-through CBR). We elaborate on the details of Cut-through CBR in the following.

For a block $B = \left[ {H,{T_1},{T_2}, \cdots ,{T_K}} \right]$, the original CBR constructs the compact block as ${B_c} = \left[ {H,{I_1},{I_2}, \cdots ,{I_K}} \right]$, where $H$ is the header of the block and ${I_k}$ is a the 6-byte short hash of transaction ${T_k}$. When relaying a compact block, instead of sending out the inv message first as in the original CBR, the sender node in our Cut-through CBR immediately broadcasts the block header $H$ to its adjacent peers. In particular, our Cut-through CBR requires the nodes to check whether the block header is valid. Specifically, after receiving the block header, the nodes need to perform the header validation by checking whether the nonce contained in $H$ fulfills (\ref{eqn:mininghash}) to validate the PoW solution. Compared with the full validation\footnote{In the original CBR for Bitcoin, the full validation of a compact block consists of the header validation and the transaction validation. The header validation checks whether the PoW is valid, while the transaction validation check whether the transactions conveyed by the compact block are valid.} of a compact block in the original CBR, this header validation is provisional. As soon as the header $H$ is received, the provisional header validation can begin before the whole compact block is received, consuming much less computation time. This provisional header verification can prevent malicious nodes from generating fake blocks at low cost. Furthermore, to maintain security, if a node in our Cut-through CBR finds an invalid transaction later during the compact block relaying, this node will suspend the compact-block relaying and alert its peers by sending an abort message, as explained in the following. 

To further elaborate our protocol, as illustrated in Fig. 1(c), we consider a scenario in which a peer P sends out $H$, and peer Q is one of P's neighbors. After peer Q receives $H$ from peer P, peers Q and P execute the following operational steps in order: 
\begin{enumerate}
\item Peer Q first checks whether the received $H$ is in its local pool. If it is, peer Q does nothing and discards $H$. If it is not, peer Q sets peer P as this block's seed peer and goes to the next operation.
\item Peer Q performs the provisional header validation of $H$ to verify the PoW solution. If the provisional header validation passes, peer Q puts $H$ into its local pool and broadcasts it to all adjacent peers (except the peer P). Furthermore, peer Q sends a query message to peer P to ask for the transaction hashes in the compact block and then goes to step 4). If the provisional header validation is not fulfilled, peer Q suspends all the operations related to this block.
\item After receiving the query message from peer Q, peer P sends peer Q the transaction hashes in chunks. Specifically, every consecutive $M$ transaction hashes are grouped into a hash chunk, and the hash chunks are sent one after another. Peer P keeps sending the hash chunks to peer Q as long as it has any of the hash chunks (either generated from peer P itself if peer P is the miner of this block or received from peer P's seed peers).
\item Peer Q receives the hash chunks sent from peer P and performs the chunk validation by validating the transactions indexed by the transaction hashes. In case peer Q lacks some of the transactions on its local storage, it requests the missing transactions from other nodes by sending the getblocktxn message, as in the original CBR \cite{corallo2017compact}. Meanwhile, peer Q also listens for query messages from other peers who treat peer Q as their seed peer of the compact block. As illustrated in Fig. \ref{fig_1}(c), peer Q receives a peer R' query message, peer Q then answers peer R with its validated hash chunks. 
\item If peer Q finds an invalid transaction during the chunk validation, it will suspend the propagation of this compact block. Furthermore, it will clear all the compact-block related messages related to the compact block from its storage and send an abort message to alert its peers that query for the same compact block. If all transactions pass validation, this block is valid, and peer Q reconstructs the block and appends it to the blockchain. 
\end{enumerate}
Our Cut-through CBR protocol processes the reception and the forwarding of a compact block's small chunks in parallel, in a pipeline manner. It thus speeds up the propagation of a huge compact block. 

\section{Coded Design: Leveraging Rateless Coding}\label{section:Coded}
In Cut-through CBR, a peer can only retrieve the compact block’s chunks from one peer. This section leverages rateless erasure codes to allow a peer to retrieve the chunks from multiple peers. This coded design incorporates fountain codes \cite{luby2002lt} (a kind of rateless erasure codes) into Cut-through CBR to efficiently utilize the upload bandwidths of multiple peers \cite{gkantsidis2005network}. 

In the coded design, instead of sending $M$ transaction hash chunks $\left\{ {{p_1},{p_2}, \cdots ,{p_M}} \right\}$, the block miner performs fountain coding over the hash chunks to generate the equal-size coded symbols $\{ {x_1},{x_2}, \cdots {x_i}, \cdots \}$ according to the coding scheme proposed in \cite{wang2007network}. Specifically, the coding scheme chooses a set of $M$ coding cofficients ${C_i} = \{ {c_1},{c_2},{c_3} \ldots ,{c_M}\}$ in a finite field. Then we have
\begin{equation}
	\label{eqn:networkCoding}
	{x_i} = \sum\limits_{j = 1}^M {{c_j}{p_j}}
\end{equation}
We refer to $\{ {x_1},{x_2} \ldots ,{x_i},...\}$ as the coded hash chunks. Then the miner sends these coded hash chunks (${C_i}$ is also sent out together with ${x_i}$) to the peers querying this compact block. After receiving a coded hash chunk, a peer relays it to some of its peers querying for the same compact block. In this way, each node can be the seed peer of a number of its peers during the compact-block relaying process. Furthermore, fountain code guarantees that the likelihood of decoding a set of coded chunks into the compact-block approaches $1$ when $n = M(1 + \varepsilon)$ coded chunks have been received, where $\varepsilon$ is typically less than $2\%$ \cite{luby2002lt}.

\begin{figure}[!t]
	\centering
	\includegraphics[width=3.5in]{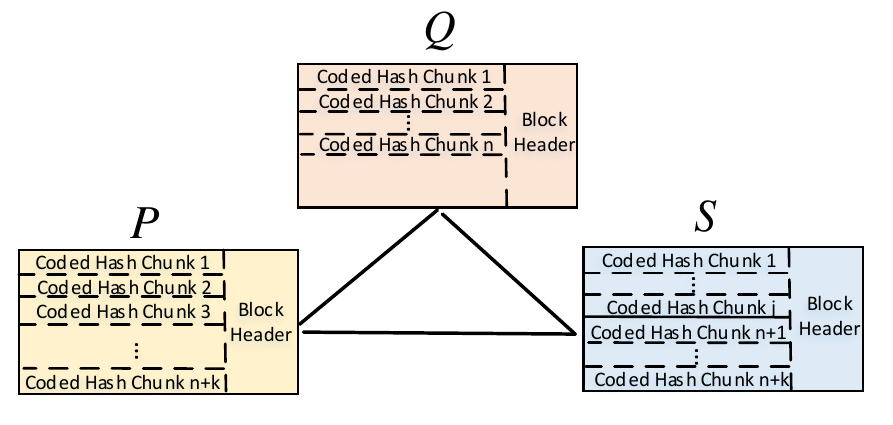}
	\caption{The network topology of three nodes (peer P, peer Q, and peer S) relaying one compact block using Coded Cut-through CBR. Node P is the miner of the compact block in this example.}
	\label{fig_2}
\end{figure}

\begin{figure}[!t]
	\centering
	\includegraphics[width=3.5in]{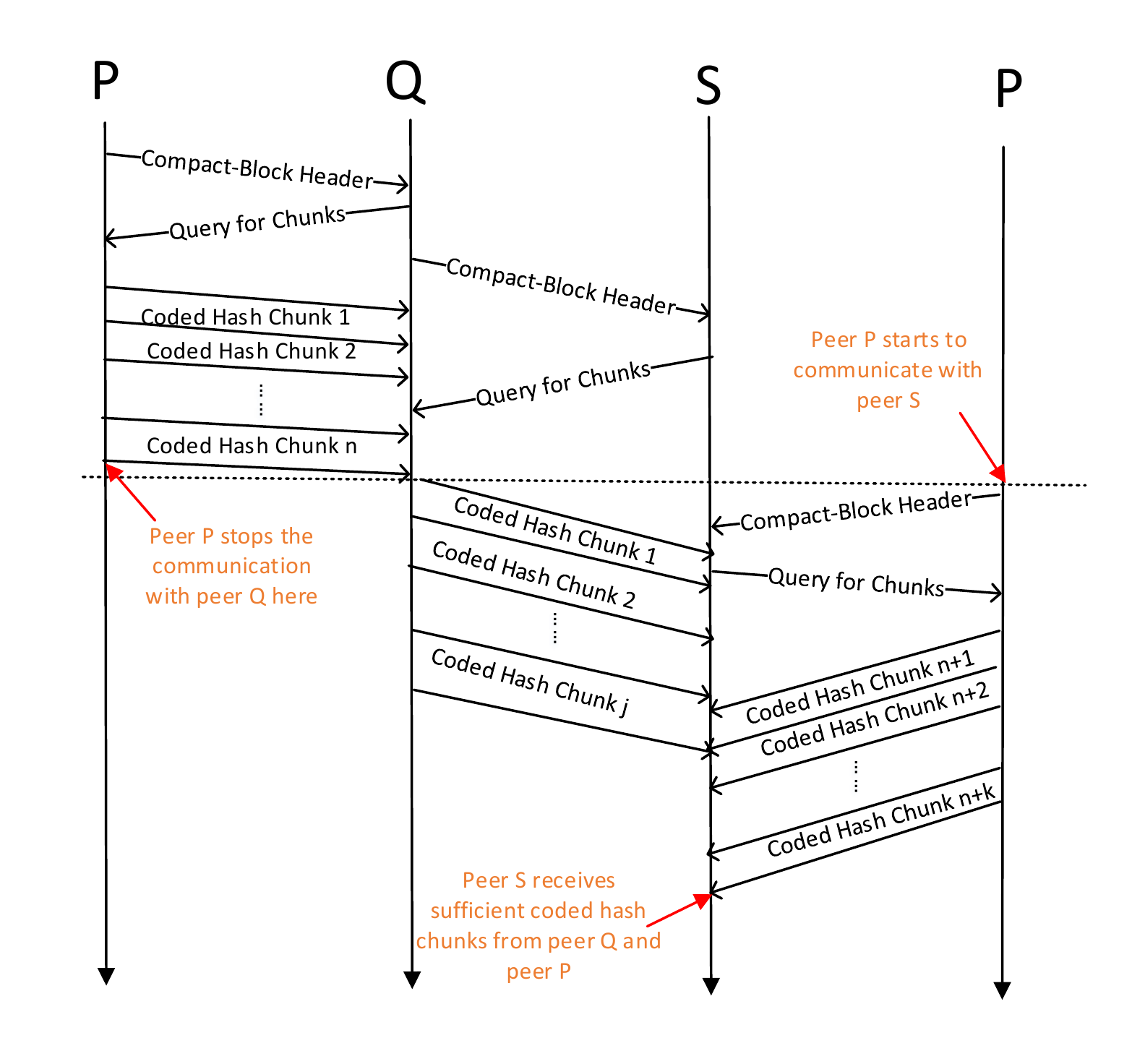}
	\caption{The illustration for Coded Cut-through CBR.}
	\label{fig_3}
\end{figure}

We name the rateless-erasure-codes enhanced protocol Coded Cut-through CBR. We describe the operational steps of Coded Cut-thought CBR using an example network topology illustrated in Fig. \ref{fig_2}, where peer P is a miner with two neighbor peers: peer Q and peer S. In particular, peer Q and peer S are also the neighbors of each other. After peer P mines a block, it sends the block header $H$ to peer Q. After peer Q receives $H$, peers Q and P execute the following operational steps in order (these operational steps are illustrated in Fig. \ref{fig_3}): 
\begin{enumerate}
	\item Peer Q first checks whether it has the block using in $H$. If it has the block, peer Q does nothing. If it does not have the block, peer Q sets peer P as this block’s seed peer and goes to step 2.
	\item Peer Q performs the provisional header validation over $H$ to validate the PoW. If the PoW is valid, peer Q puts $H$ into its local pool and relays $H$ to its neighbor peer (in this case, its neighbor peer is peer S). After that, peer Q sends a query message to peer P querying the transaction hashes. If the PoW is not valid, peer Q aborts and does not relay this compact block. 
	\item After receiving the query message from peer Q, peer P sends peer Q the coded hash chunks generated by fountain coding. Specifically, peer P keeps generating and sending the coded hash chunks ${x_i},i = 1,2, \cdots$ to peer Q until peer Q sends back an ACK message announcing that it receives sufficient coded hash chunks to decode the fountain code. In Fig. 3, peer P sends a total of $n$ coded hash chunks to peer Q. 
	\item When receiving coded hash chunks ${x_i},1 \le i \le n$ from peer P, peer Q also relays its received coded hash chunks to peer S that queries the chunks after receiving $H$ from peer Q. 
	\item After receiving the ACK, peer P stops sending its coded hash chunks to peer Q. Peer P then sends $H$ to peer S seeking to be a potential seed peer of peer S. Once receiving the query message sent from peer S, peer P starts generating and sending coded hash chunk ${x_i},i \ge n + 1$ to peer S. 
	\item At this time, peer S has two seed peers of this compact block, namely, peer P and peer Q. Peer S keeps receiving the coded hash chunks sent from either peer P or peer Q. Once it gets sufficient coded hash chunks for decoding the compact block, peer S sends the ACK to peer P and peer Q one after another. In Fig. \ref{fig_3}, peer S receives $j,j \le n$ coded hash chunks from peer Q and $k,k \le n^{'}$ coded hash chunks from peer P. Here, for peer S, we have $j + k = n^{'}$, $n^{'}$ is the number of coded hash chunks required to decode the compact block.
\end{enumerate}
After receiving sufficient coded chunks, peer Q and peer S decode the fountain-code and validate all transactions. If all transactions pass validation, each peer reconstructs this block and appends it to the blockchain. 

In Fig. \ref{fig_3}, if the peer Q’s upload bandwidth is much lower than that of other peers, then we have $j \ll n$. In this case, if using standard relaying (SR) (see Section 3.4) and Cut-through CBR, peer S can do nothing except receiving the information sent from peer Q. In Coded Cut-through CBR, however, the available upload bandwidth of other nodes (in this case, the available upload bandwidth of peer P) can be further used to relay the compact block to peer S. Hence, Coded Cut-through CBR leverages the P2P network topology to use the upload bandwidths of all peers efficiently. Thus, it can speed up the information propagation in the Bitcoin blockchain. We theoretically analyze our Cut-through CBR and Coded Cut-through CBR protocols to validate their potential gains in the next section.

\section{Performance Analysis}
This section presents the effect of the block propagation delay on the Bitcoin blockchain's TPS and the analysis of the block propagation latencies of the four propagation protocols (SR, CBR, Cut-through CBR and Coded Cut-through CBR).
\label{section:analysis}
\subsection{The effect of the block propagation delay on TPS}
We first calculate the TPS of the Bitcoin blockchain. We denote the block size by ${S_{block}}$, the average transaction size by ${S_{transaction}}$, and the inter-block generation time by ${T_B}$. The TPS of the blockchain is calculated as  
\begin{equation}
	\label{eqn:TPSCal}
	TPS = \frac{{{S_{block}}}}{{{S_{transaction}}{T_B}}}
\end{equation}
In Bitcoin, we have ${S_{block}} = 1$ MB, ${S_{transaction}} = 380.04$ B and ${T_B} = 600$ s, and thus the TPS of Bitcoin is 
$TP{S_{Bitcoin}} = \frac{{1048576}}{{380.04*600}} \approx 4.6$. 

According to (\ref{eqn:TPSCal}), a simple way to increase the TPS is to increase ${S_{block}}$ or to decrease ${T_B}$. However, merely increasing ${S_{block}}$ or decreasing ${T_B}$ cannot increase the TPS, as explained below.  It takes a certain time to propagate a block over the p2p network of the blockchain. Let $L$ denote the time latency of propagating a block from the miner to almost all nodes in the whole network (e.g., $90\%$ of the nodes in the network). First, the increase of ${S_{block}}$ leads to an increase of $L$, which, in turn, may compromise the security of the blockchain \cite{gervais2016security}. In particular, $L$ should be sufficiently smaller than ${T_B}$. The closer $L$ gets to ${T_B}$, the more forks, more orphan blocks, and more chain re-organizations there will be. According to \cite{klarman2018bloxroute}, the probability for a fork to occur at another miner is approximately 
\begin{equation}
	\label{eqn:forkCal}
	P(fork|{T_B} = 600) = 1 - {e^{\frac{{ - L}}{{{T_B}}}}}
\end{equation}
In the extreme case, the system will be exposed to security vulnerabilities such as double-spend attacks\cite{gervais2016security}. Motivated by (\ref{eqn:forkCal}), this paper uses a performance metric, propagation divergence factor, to quantify the effect of block propagation latency. \\
\textsl{ \textbf{Propagation Divergence Factor}}: The propagation divergence factor is defined as
\begin{equation}
	\label{eqn:PDF}
	\Delta  = \frac{L}{{{T_B}}}
\end{equation}
where $\Delta  \ge 0$  since $L \ge 0$. The larger $\Delta $, the larger the divergence (e.g., asynchronies and discrepancies) is among the different local replicas of the blockchain maintained by different nodes. In the extreme case where $L = 0$, we have $\Delta  = 0$ meaning that the blockchains of all the nodes in the network become instantaneously synchronized. 

The main purpose of our protocols is to suppress the increase of $L$ when we increase ${S_{block}}$. In this way, we can boost the TPS of the Bitcoin blockchain while maintaining the security of the blockchain. According to (\ref{eqn:TPSCal}), increasing ${S_{block}}$ by 100x can increase the TPS by 100x, yielding $TP{S_{Bitcoin}} \approx 460$. In the next subsection, we analyze the block propagation latency $L$ caused by different block propagation protocols. 

\begin{figure}[!t]
	\centering
	\includegraphics[width=3.5in]{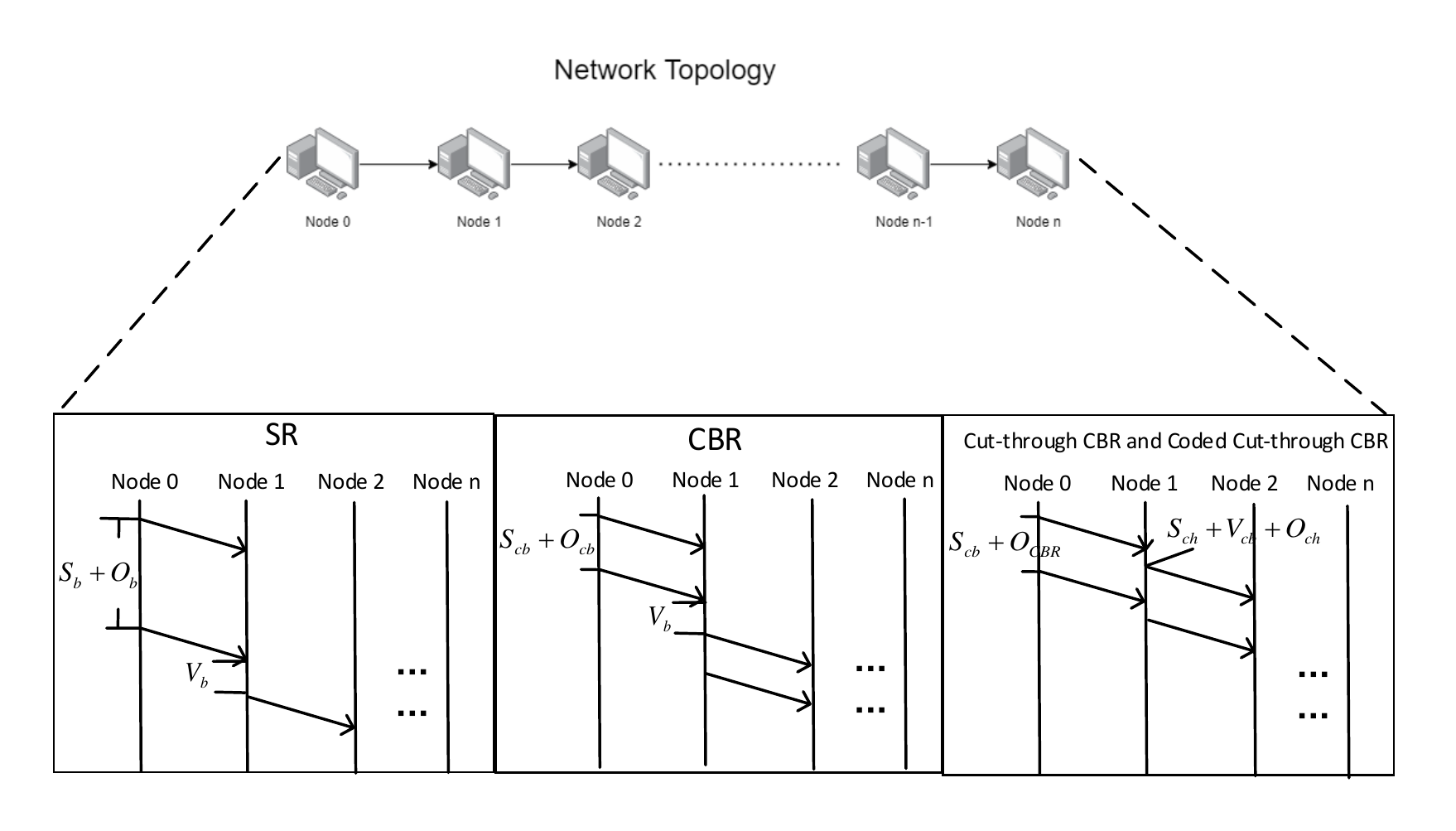}
	\caption{A simple linear network with $n+1$ nodes.}
	\label{fig_4}
\end{figure}

\subsection{The comparison of the block propagation latency using different protocols}
We next analyze the block propagation latencies of the four propagation protocols (SR, CBR, Cut-through CBR and Coded Cut-through CBR). For simple illustration, we consider a network with a linear topology consisting of 
$n + 1$ nodes as shown in Fig. \ref{fig_4}. We assume the communication bandwidth of each node is $B$. We calculate the block propagation latency of propagating a block from node $0$ to node $n$ in this linear network.

Let ${S_{ip}}$ be the IP packet's payload size and ${S_h}$ be the IP packet header size. When using SR to propagate blocks, with IP packet fragmentation, a block needs to be cut into $\frac{{{S_{block}}}}{{{S_{ip}}}}$ IP packets. The block propagation latency from node $0$ to node $n$ caused under the SR protocol is given by 
\begin{equation}
	\label{eqn:L_SR}
	{L_{SR}} = \frac{{n({S_{block}} + {O_b})}}{B} + n{V_b}
\end{equation}
where ${O_b} = \frac{{{S_{block}}}}{{{S_{ip}}}}{S_h}$ is the overhead caused by the IP packet fragmentation when relaying a block over one hop between two neighbor nodes, and ${V_b}$ is the time incurred by a full-block validation process.

Let ${N_t}$ be the number of transactions in a block. The size of a compact block is given by
\begin{equation}
	\label{eqn:Size_CBR}
	{S_{cb}} = {N_t}{S_I} + {S_{h\_cb}}
\end{equation}
where ${S_I}$ is the size of a short transaction hash and ${S_{h\_cb}}$ is the size of the compact-block header. When using CBR, a compact block needs to be cut into $\frac{{{S_{cb}}}}{{{S_{ip}}}}$ IP packets. Recall that in Bitcoin, nodes have 10 min (${T_B} = 10$ min) to propagate their collected transactions to all other nodes before the transactions are announced in the latest block. For the sake of simplicity, we assume that all transactions within the block are already available at the nodes given the long duration of 10 min, and there is no need to go and fetch the transactions. In fact, \cite{corallo2017compact} found that in the Bitcoin network, before a new block is found, transactions obtained by all nodes are almost synchronized. Then, the block propagation latency from node $0$ to node $n$ under CBR is given by
\begin{equation}
	\label{eqn:L_CBR}
	{L_{CBR}} = \frac{{n({S_{cb}} + {O_{cb}})}}{B} + n{V_b}
\end{equation}
where ${O_{cb}} = \frac{{{S_{cb}}}}{{{S_{ip}}}}{S_h}$ is the overhead caused by the IP packet fragment when relaying a compact block over one hop. 

From (\ref{eqn:L_SR}) and (\ref{eqn:L_CBR}), we can see that CBR speeds up the block propagation significantly, since ${S_{cb}} \ll {S_b}$. In particular, if ${V_b}$ is negligible, CBR can reduce the block propagation latency by $\frac{{{S_{block}}}}{{{S_{cb}}}}$ times compared with SR. However, ${L_{CBR}}$ is still proportional to $n{S_{cb}}$. Consequently, the increase of ${S_{cb}}$ can lead a significant increase of ${L_{CBR}}$, making it impossible to increase TPS by inserting more transaction hashes into each compact block.  

We next analyze the propagation latency of the cut-through forwarding scheme used in both Cut-through CBR and Coded Cut-through CBR. Let $k \ge 1$ be the number of chunks per compact block and ${S_{ch}}$ be the chunk size. We thus have ${S_{cb}} = k{S_{ch}}$. Let ${V_{ch}} = \frac{{{V_b}}}{k}$ be the time needed to validate the transactions in a chunk and ${O_{ch}}$ be the overhead of IP packet fragmentation when relaying a chunk over one hop. The block propagation latency from node $0$ to node $n$ under Cut-through CBR is given by
\begin{equation}
	\label{eqn:L_CT}
	{L_{CT - CBR}} = k(\frac{{{S_{ch}} + {O_{ch}}}}{B} + {V_{ch}}) + (n - 1)(\frac{{{S_{ch}} + {O_{ch}}}}{B} + {V_{ch}})
\end{equation}
The first term of (\ref{eqn:L_CT}) is the block propagation latency in the first hop (from node $0$ to node $n$) and the second term of (\ref{eqn:L_CT}) is the block propagation latency over the rest of the hops (from node $0$ to node $n$). We further investigate the effect of $k$ when using the cut-through forwarding scheme in the following two cases: 

If $1 \le k \le \frac{{{S_{cb}}}}{{{S_{ip}}}}$, then we have $\frac{{{S_{cb}}}}{k} = {S_{ch}} \ge {S_{ip}}$, which means that we need to perform IP packet fragmentation over each chunk. Specifically, a chunk needs to be cut into $\frac{{{S_{ch}}}}{{{S_{ip}}}}$ IP packets and ${O_{ch}} = \frac{{{S_{ch}}}}{{{S_{ip}}}}{S_h} = \frac{{{O_{cb}}}}{k}$. From (\ref{eqn:L_CT}), we then have 
\begin{equation}
	\label{eqn:L_CT-1}
	{L_{CT - CBR}} = (\frac{{{S_{cb}} + {O_{cb}}}}{B} + {V_b}) + \frac{{(n - 1)}}{k}(\frac{{{S_{cb}} + {O_{cb}}}}{B} + {V_b})
\end{equation}
Compared with (\ref{eqn:L_CBR}), the increase of $k$ in (\ref{eqn:L_CT-1}) will significantly reduce the block propagation latency. We denote the gain obtained from the cut-through forwarding scheme over the original CBR by ${g_{CT - CBR}} = {L_{CBR}} - {L_{CT - CBR}}$. The gain ${g_{CT - CBR}}$ can be computed as
\begin{equation}
	\label{eqn:g_CT-1}
	\begin{array}{l}
	{g_{CT - CBR}} = \frac{{n({S_{cb}} + {O_{cb}})}}{B} + n{V_b} - (k + n - 1)(\frac{{{S_{ch}} + {O_{ch}}}}{B} + {V_{ch}})\\
	= \frac{{n({S_{cb}} + {O_{cb}}) - (k + n - 1)(\frac{{{S_{cb}}}}{k} + \frac{{{O_{cb}}}}{k})}}{B} + n{V_b} - (k + n - 1)\frac{{{V_b}}}{k}\\
	= (1 - \frac{1}{k})(n - 1)(\frac{{({S_{cb}} + {O_{cb}})}}{B} + {V_b})
	\end{array}
\end{equation}
Note that since $k \ge 1$ and $1 - \frac{1}{k} > 0$, we have ${g_{CT - CBR}} > 0$. That is, the cut-through forwarding scheme outperforms the traditional CBR when $1 \le k \le \frac{{{S_{cb}}}}{{{S_{ip}}}}$. Eqn. (\ref{eqn:L_CBR}) and Eqn. (\ref{eqn:g_CT-1}) lead to several conclusions:
\begin{enumerate}
	\item When $k = 1$ or $n = 1$, both Cut-through CBR and Coded Cut-through CB downgrade to CBR. 
	\item When $k$ is fixed, the gain obtained from the cut-through forwarding scheme becomes larger with the increase of ${S_{cb}}$. The cut-through forwarding scheme hence is expected to perform well even if ${S_{cb}}$ is very large.
	\item When ${S_{cb}}$ is fixed, ${L_{CT - CBR}}$ is a decreasing function of $k$. The increase of $k$ achieves lower propagation latency. 
\end{enumerate}

If $k > \frac{{{S_{cb}}}}{{{S_{ip}}}}$, then we have $\frac{{{S_{cb}}}}{k} = {S_{ch}} < {S_{ip}}$. In this case, we use one IP packet to transmit one chunk with the IP packet size smaller than the maximum IP packet size. Hence ${O_{ch}} = {S_h}$ and the gain ${g_{CT - CBR}}$ is given by 
\begin{equation}
	\label{eqn:g_CT-2}
	\begin{array}{l}
		{g_{CT - CBR}} = \frac{{n({S_{cb}} + {O_{cb}})}}{B} + n{V_b} - (k + n - 1)(\frac{{{S_{ch}} + {O_{ch}}}}{B} + {V_{ch}})\\
		= \frac{{n({S_{cb}} + {O_{cb}}) - (k + n - 1)(\frac{{{S_{cb}}}}{k} + {S_h})}}{B} + n{V_b} - (k + n - 1)\frac{{{V_b}}}{k}
	\end{array}
\end{equation}
From (\ref{eqn:g_CT-2}), we can see that ${L_{CT - CBR}}$ is no more a decreasing function of $k$. To find the optimal k that maximizes the gain expressed in (\ref{eqn:g_CT-2}), we can set the derivative of  ${L_{CT - CBR}}$ with respect to $k$ to zero: $\frac{{d{g_{CT - CBR}}}}{{dk}} = 0$, and solve the equation. The optimal k is given by
\begin{equation}
	\label{eqn:choice_k}
	{k^*} = \sqrt {\frac{{B(n - 1)({V_b} + {S_{cb}})}}{{{S_h}}}}
\end{equation}
where the gain ${L_{CT - CBR}}$ achieves its maximum.

The above analysis for a linear network shows the potential gain brought by our Cut-through CBR protocol. In the next section, we will investigate the performances of our Cut-through CBR and Coded Cut-through CBR protocols under a practical network setup. 

\section{Simulation Results}
\label{section:simulation}
This section presents simulation results comparing the performance of the four block-propagation protocols: SR, CBR, Cut-through CBR, and Coded Cut-through CBR. Specifically, using the network simulator 3 (NS3) \cite{henderson2008network}, we investigate the block-propagation latencies of these four propagation protocols in a simulated Bitcoin network. 

\subsection{Simulation Setup}
Our investigation is performed on a simulator extended from the Bitcoin simulator implemented by \cite{gervais2016security}. The Bitcoin simulator, written in NS3, is designed to simulate Bitcoin nodes' behavior in the Bitcoin network. The Bitcoin simulator can simulate a Bitcoin network with thousands of nodes. The Bitcoin simulator retrieved the current geographical node distribution from bitnodes.21.co and adopted the distribution to define its simulated nodes' locations. In the simulated Bitcoin network, the connection between two nodes is established as a point-to-point (P2P) link. Each P2P link has a random bandwidth and transmission delay (according to the geographical location) following a statistical distribution from Verizon \cite{verizon} and testmy.net \cite{testmy}. The parameters in Table I presents the parameters used in the Bitcoin simulator \cite{gervais2016security}.

\begin{table}[]
	\caption{PARAMETERS OF THE BITCOIN SIMULATOR}
	\label{tab:my-table}
	\centering
	\begin{tabular}{|l|l|}
		\hline
		Parameter                   & Bitcoin Simulator                                                                           \\ \hline
		\# of the nodes             & 6000                                                                                        \\ \hline
		Block interval              & 10 min                                                                                      \\ \hline
		\# of the connection        & \begin{tabular}[c]{@{}l@{}}Distribution according to\\ Miller et al. \cite{miller2015discovering}\end{tabular} \\ \hline
		Geographical distribution   & \begin{tabular}[c]{@{}l@{}}Distribution according to \\ actual blockchains\end{tabular}     \\ \hline
		Bandwidth propagation delay & \begin{tabular}[c]{@{}l@{}}6 regional bandwidth and \\propagation delay\end{tabular}        \\ \hline
	\end{tabular}
\end{table}
The original Bitcoin simulator already included the SR protocol. We implemented CBR, Cut-through CBR and Coded Cut-through CBR in the simulator. The extended simulator allows us to set parameters such as the block size and the chunks per block for Cut-through CBR and Coded Cut-through CBR. We performed simulations to investigate the effect of block size on the network block propagation latency and the propagation divergence factor defined in (\ref{eqn:PDF}) under different block propagation protocols. Our overall simulation experiment consists of several simulation runs. We fixed a block size for each run and then ran separate simulations to evaluate SR, CBR, Cut-through CBR, and Coded Cut-through CBR. Each simulation run, for a specific relay protocol, monitored the statistics of up to $10\,000$ blocks created during the simulation to smooth the experimental results.

\begin{figure}[!t]
	\centering
	\includegraphics[width=3.5in]{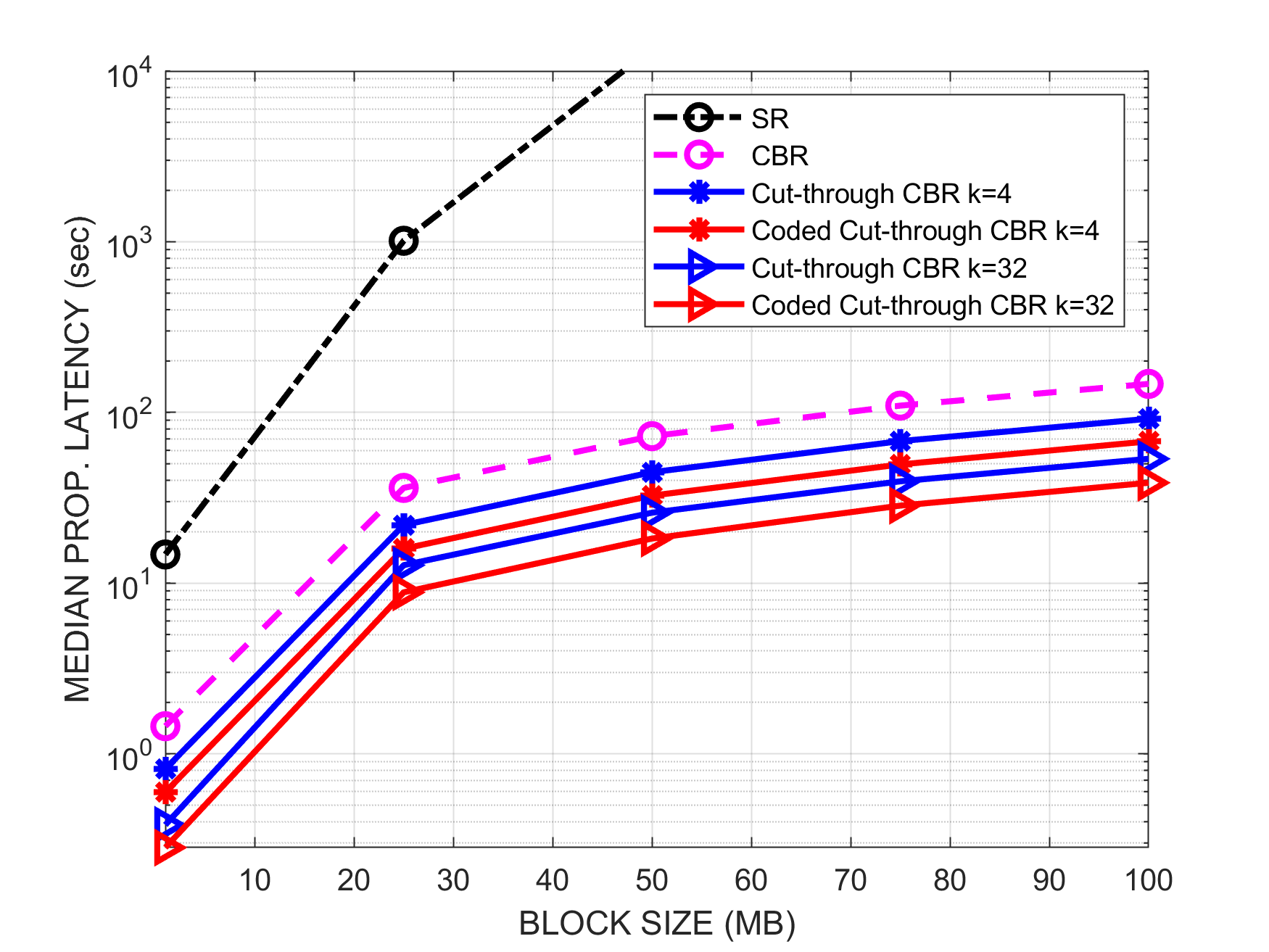}
	\caption{Median block propagation latency versus block size for different block propagation protocols.}
	\label{fig_5}
\end{figure}

\begin{figure}
	\begin{subfigure}{.5\textwidth}
		\centering
		\includegraphics[width=.8\linewidth]{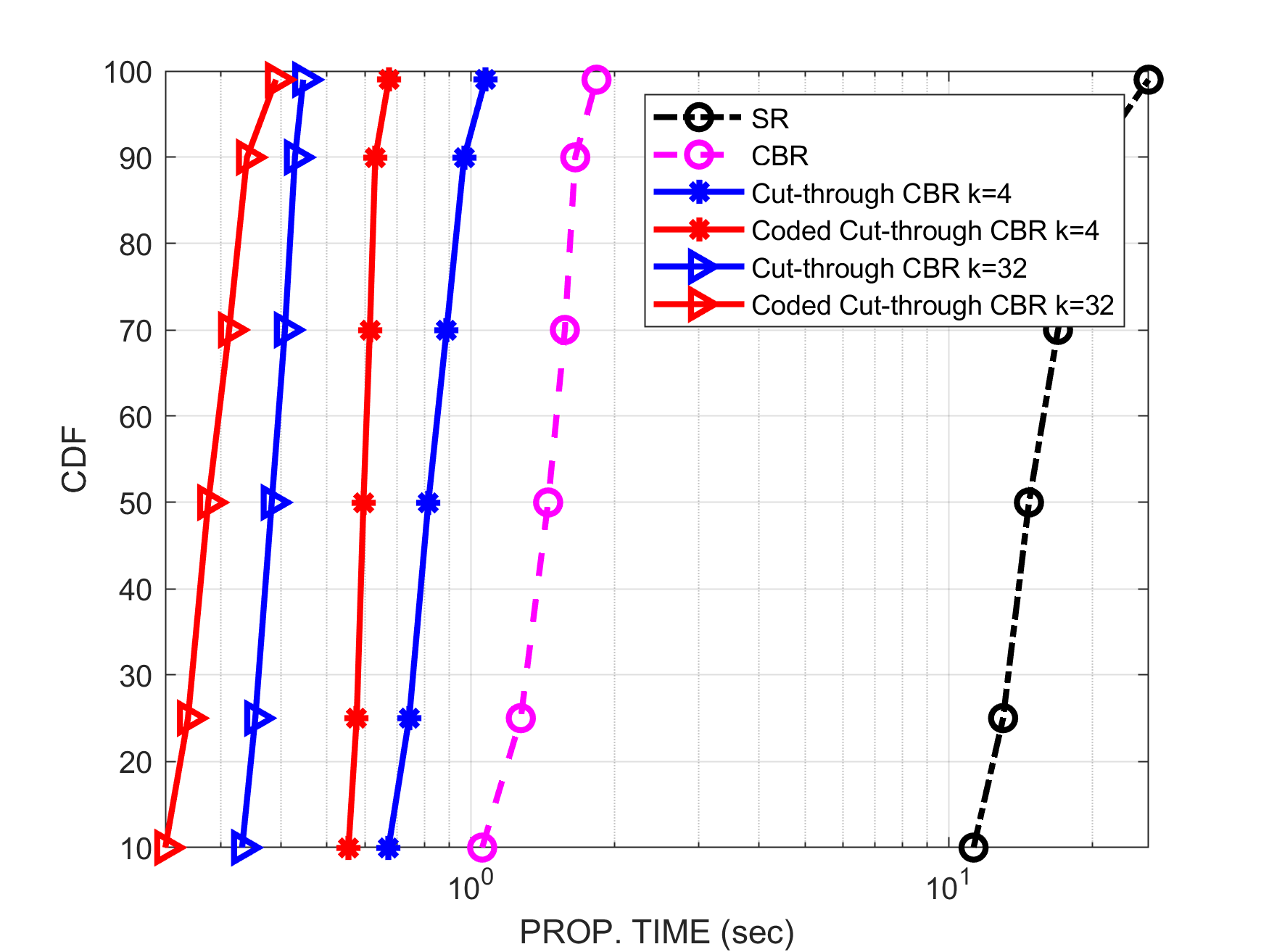}  
		\caption{BLOCK SIZE: 1 MB}
		
	\end{subfigure}
	\begin{subfigure}{.5\textwidth}
		\centering
		\includegraphics[width=.8\linewidth]{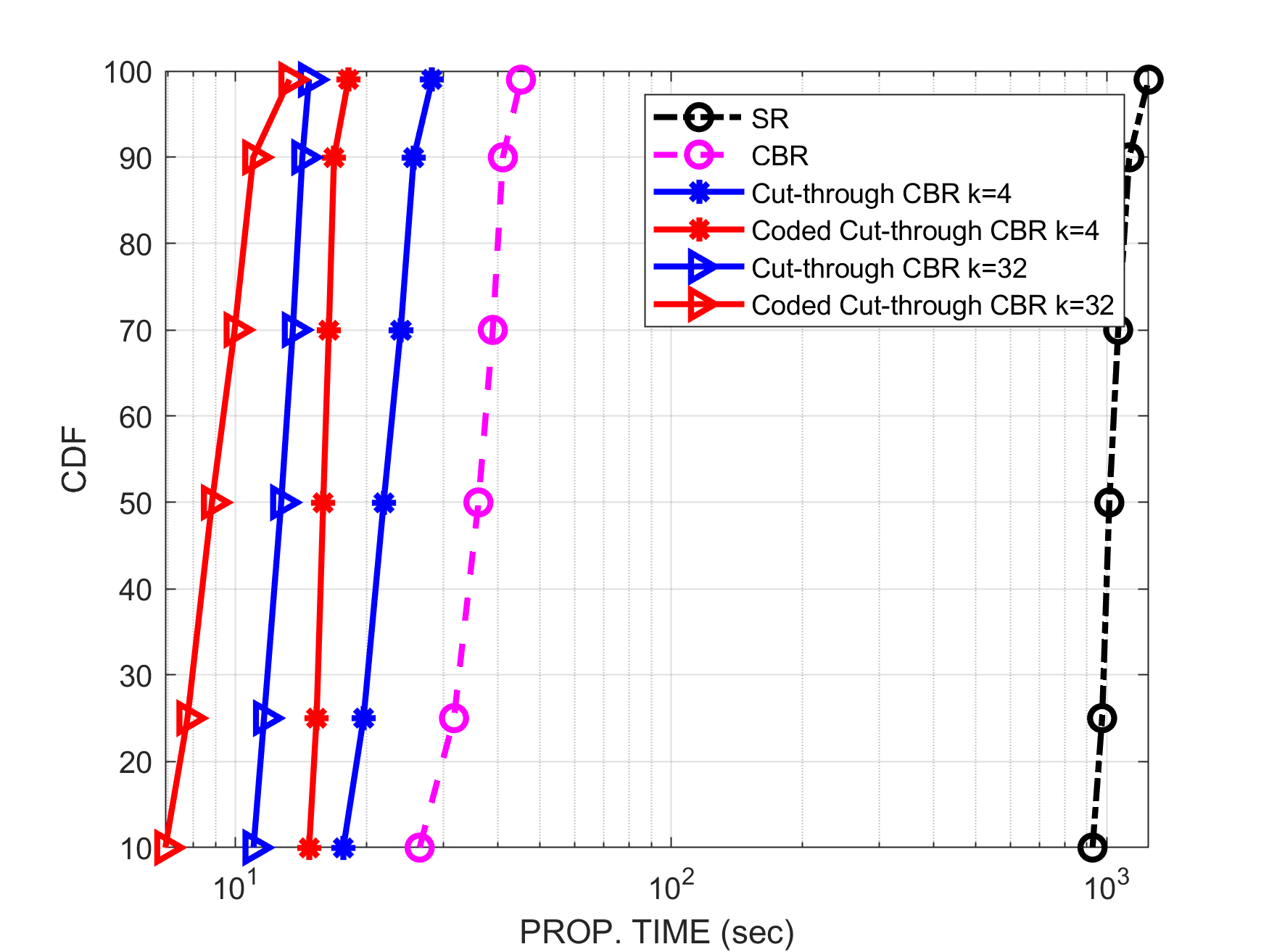} \caption{BLOCK SIZE: 25 MB}
	\end{subfigure}
	
	\begin{subfigure}{.5\textwidth}
		\centering
		\includegraphics[width=.8\linewidth]{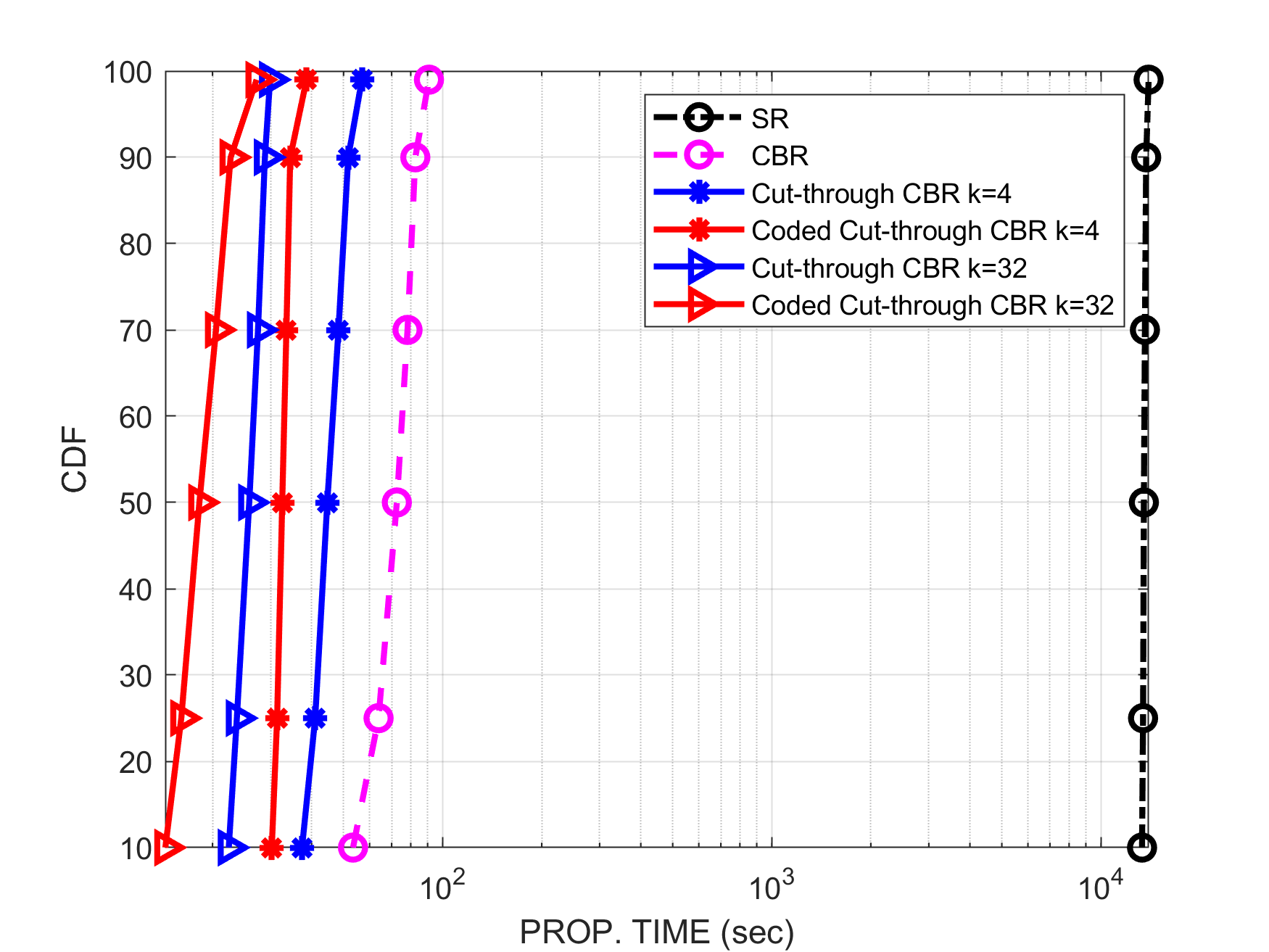} \caption{BLOCK SIZE: 50 MB}
		
	\end{subfigure}

	\begin{subfigure}{.5\textwidth}
		\centering
		\includegraphics[width=.8\linewidth]{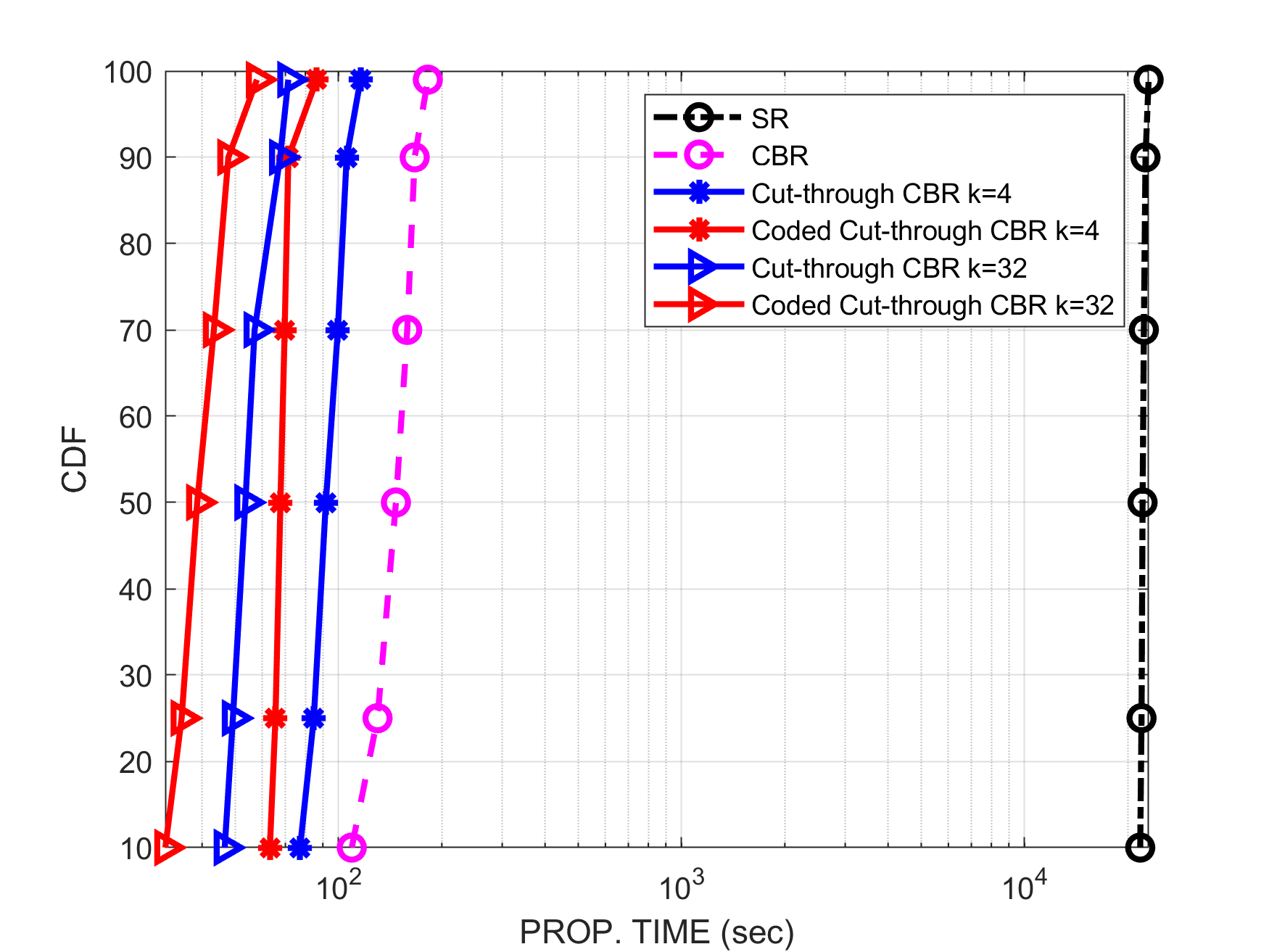}\caption{BLOCK SIZE: 100 MB}
	\end{subfigure}
	\caption{CDF of block propagation latency for different block propagation protocols.}
	\label{fig_6}
\end{figure}

\subsection{Experimental Results}
Fig. \ref{fig_5} and Fig. \ref{fig_6} show the median and the CDF of the block propagation latency, respectively. We have the following observations:
\begin{enumerate}
	\item When relaying a large block, cut-through forwarding can significantly speed up block propagation with respect to SR and CBR. For example, Fig. 5 and Fig. 6 show that when relaying a block larger than $25$ MB, Cut-through CBR and Coded Cut-through CBR reduce both the median and the tail of block propagation latency by up to three times compared with CBR and by up to more than one hundred times compared with SR.  
	\item Coded Cut-thought CBR can further speed up the block propagation compared to Cut-through CBR. For example, Fig. 5 and Fig. 6 show that the median and the tail of block propagation latency of Coded Cut-through CBR are both up to two times smaller than that of cut-through CBR for different $k$. 
\end{enumerate}

\begin{figure}[!t]
	\centering
	\includegraphics[width=3.5in]{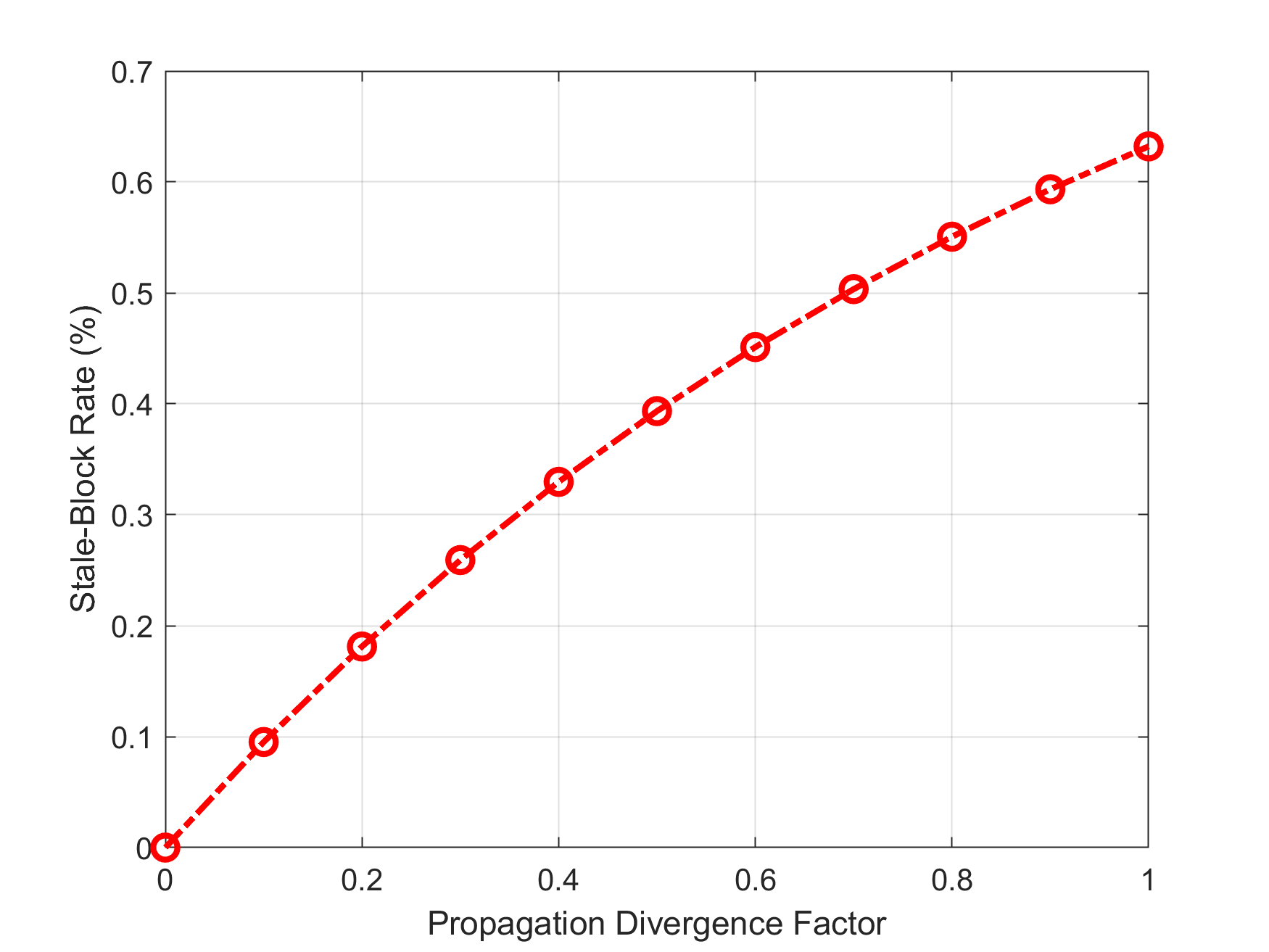}
	\caption{Stale-block size versus propagation divergence factor derived using (\ref{eqn:forkCal}).}
	\label{fig_7}
\end{figure}

\begin{figure}[!t]
	\centering
	\includegraphics[width=3.5in]{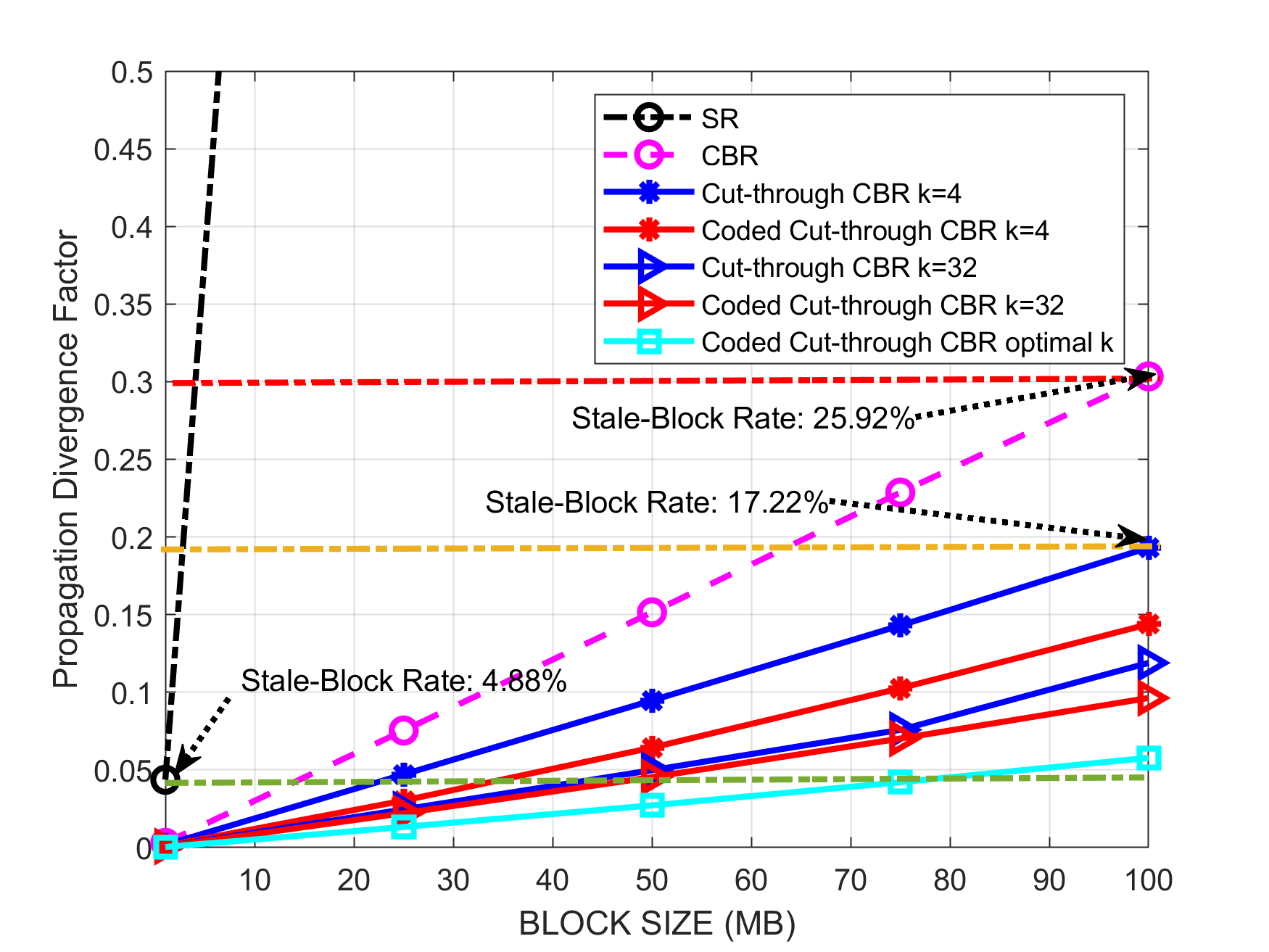}
	\caption{Propagation divergence factor versus block size for different block propagation protocols.}
	\label{fig_8}
\end{figure}

Fig. \ref{fig_7} shows the stale-block rate versus the propagation divergence factor $\Delta $. We can see that large $\Delta $ increases stale-block rate significantly. The reason is that when the block propagation divergence factor is large, a miner cannot receive the latest announced block (mined by others) in time and may still announce its mined block with the same block height, leading to a large stale-block rate. Fig. 8 shows the propagation divergence factor $\Delta $ of different protocols when different block sizes are used. From Fig. \ref{fig_8}, we can see that the propagation divergence factors $\Delta $ of SR and CBR increase significantly when the block size is large, leading to a large stale-block rate. For example, increasing block size from 1 MB to 100 MB (100x) increases the propagation divergence factor $\Delta $ by 30x and increases the stale-block rate by around 30x for CBR. However, Cut-through CBR and Coded Cut-through CBR can suppress the increase of $\Delta $ with the increase of block size. In particular, the $\Delta $ of Coded Cut-through CBR with the optimal $k$ (the optimal $k$ is obtained through experiments), when the block size is 100 MB, is roughly equal to the $\Delta $ of SR when the block size is 1MB.

In summary, Cut-through CBR and Coded Cut-through CBR can be used for propagating 100 MB blocks while maintaining the propagation latency to the level of the traditional block propagation protocol that propagates 1 MB blocks, hence increasing the TPS capacity by 100x. 

\section{Conclusion}
\label{section:conclusion}
We proposed a new blockchain networking protocol  to increase the TPS of the Bitcoin blockchain. When a large block size is used in the Bitcoin blockchain, original block propagation protocols, such as SR and CBR, suffer from large block propagation delays to large stale-block rate, compromising the blockchain's security. In this work, we put forth a two-pronged approach to increase the block size without inducing extra propagation latency. First, we design a Cut-through CBR that enables parallel reception and forwarding of compact blocks at relay nodes. Second, we design a Coded Cut-through CBR that incorporates rateless erasure codes into Cut-through CBR to further increase efficiency. Our simulation results demonstrate that our protocols can significantly reduce the block propagation latency and suppress the stale-block rate. Specifically, our protocols can increase the TPS of the Bitcoin blockchain by 100x without compromising the blockchain's security. More importantly, our approach only needs to rework the communication and networking architecture of the current Bitcoin blockchain without changing the data structures and crypto-functional components in them. Therefore, our protocols can be seamlessly incorporated into the existing Bitcoin blockchain. The implementation of our protocols in Bitcoin-like blockchains may allow the blockchains to be used in many use cases not possible currently. 
\ifCLASSOPTIONcaptionsoff
  \newpage
\fi



%



\bibliographystyle{IEEEtran}

\bibliography{database}

\end{document}